\documentclass[oldversion]{aa}
\usepackage{epsfig}
\usepackage{natbib}
\usepackage{lscape}
\bibpunct{(}{)}{;}{a}{}{,}
\begin{document}

\title{On the Age of Stars Harboring Transiting Planets\thanks{Based on 
observations collected
at the La Silla Parana Observatory,
ESO (Chile) with the CORALIE spectrograph at the 1.2-m Euler Swiss telescope and 
UVES at the VLT/UT2 Kueyen telescope
(Observing run 075.C-0185).}}

 \author{
 	    C. Melo\inst{1} \and
	    N. C. Santos\inst{2,3,4} \and
	    F. Pont\inst{3} \and
	    T. Guillot\inst{5}
	    G. Israelian\inst{6} \and
	    M. Mayor\inst{3} \and
	    D. Queloz\inst{3} \and
            S. Udry\inst{3}
   }

   \institute{
	     European Southern Observatory, 
	     Casilla 19001, Santiago 19, Chile
	 \and    
             Centro de Astronomia e Astrof{\'\i}sica da Universidade de Lisboa,
             Observat\'orio Astron\'omico de Lisboa, Tapada da Ajuda, 1349-018
             Lisboa, Portugal
	 \and
             Observatoire de Gen\`eve, 
	     51 ch.  des Maillettes, CH--1290 Sauverny, Switzerland
	 \and
	     Centro de Geof\'{\i}sica de Evora, Rua Romeo Ramalho 59, 7002-554 Evora, Portugal
         \and
	     Laboratoire Cassiop\'ee, CNRS UMR 6202, Observatoire de la C\^ote 
             d'Azur, Nice, France
	 \and
             Instituto de Astrof{\'\i}sica de Canarias, 
	     E-38200 La Laguna, Tenerife, Spain
}

   \date{Received XXX; accepted XXX}
 \abstract
   {Results of photometric surveys have brought to light the existence of
   a population of giant planets orbiting their host stars even closer than the 
hot Jupiters (HJ),
   with orbital periods below 3 days. The reason why radial velocity surveys
   were not able to detect these very-hot Jupiters (VHJ) is under discussion.
   A possible explanation is that these close-in planets are short-lived, being
   evaporated on short time-scales due to UV flux of their host stars. In this 
case, 
   stars hosting transiting VHJ planets would be systematically
   younger than those in the radial velocity sample.
   We have used the UVES spectrograph (VLT-UT2 telescope) to obtain high 
resolution 
    spectra of 5 faint stars hosting transiting planets, namely, OGLE-TR-10, 
    56, 111, 113 and TrES-1. Previously obtained CORALIE spectra of HD189733, 
and published data on
    the other transiting planet-hosts were also used. The immediate objective is 
to 
    estimate ages via Li abundances, using the Ca~II activity-age relation, and 
from
    the analysis of the stellar rotational velocity.
   For the stars for which we have spectra, Li abundances were computed as in 
   Israelian et al. (2004) using the stellar parameters derived in Santos et al. (2006). 
   The chromospheric activity index $S_{US}$ was built as the ratio of the flux 
within the core of
   the Ca II H \& K lines and the flux in two nearby continuum regions. 
   The index $S_{US}$ was
   calibrated to Mount Wilson index $S_{MW}$ allowing the computation of the 
   Ca II H \& K corrected for the photospheric contribution. These values were 
then used to 
   derive the ages by means of the Henry et al. (1996) activity-age relation.
   Bearing in mind the limitations of the ages derived by Li abundances, chromospheric 
activity,
   and stellar rotational velocities,
   none of the stars studied in this paper seem to be younger than 0.5 Gyr.

   }

   \authorrunning{Melo et al.}
   \maketitle

\section{Introduction}
%

Since the discovery of 51Peg b by \cite{Mayor-1995} the number of new extra- solar planets  has been rapidly
growing. More than about 160 extra-solar planets\footnote{For a  continuously  updated list see table at
http://obswww.unige.ch/Exoplanets} have been found and  yet  none of them look like our paradigm for
planetary formation, i.e., the Solar  System.  On the contrary, these new discoveries provide evidence that a large
variety of orbital  characteristics probably related to their formation process are actually  possible. 

Due to the considerable number of giant planets found orbiting their host stars  at short  periods (typically
below 10 days) it was soon realized that migration has to  play a key  role in their formation, since a solid
core massive enough to accrete gas is not  able to grow at such a short distance from the parent star 
\citep{Pollack-1996}.%
With
the increasing number of discoveries, statistical trends are arising,  revealing  interesting imprints thought to
be left by the migration mechanism which brought  these  planets close-in \citep{Udry-2003}. An annoying point
when dealing with  migration is  how to stop it. The pile-up of planets with $P_{orb}\sim3-3.5$ days in the the
orbital period distribution was understood  as an observational clue indicating the migration parking point.
Although appealing, this interpretation was severely challenged by the discovery  of the  transiting planets.
Some of these new transiting planets have more than a  Jupiter mass  but nevertheless they revolve around their
host stars in an even shorter orbital  period  than the so-called Hot Jupiters (HJ), typically $P_{orb}\la1-2$
days. For this  reason,  they have been called Very Hot Jupiters (VHJ).

Why these planets have not been detected more often by the radial velocity  surveys is a  matter of debate.
\cite{Gaudi-2005} advocate that the non-detection of VHJ by  the radial  velocity surveys is a statistical
effect due to the much lower frequency of VHJ  as compared  to that of HJ. 

Given their proximity to their host stars, the VHJ may experience some degree of  evaporation  due to the
heating by stellar UV photons. 
Vidal-Madjar et al. (2004) showed that HD209458b is indeed evaporating, but due to saturation of the spectral
lines, only an upper limit on the mass loss of 10$^{10}\rm\,g\,s^{-1}$ can be derived. According to theoretical
calculations, the extreme case of an efficient heating of the planet atmosphere may lead to an energy-limited
evaporation $~10^{12}\rm\,g\,s^{-1}$ for HD209458b (Lammer et al. 2003), and hence to a very efficient
evaporation, possibly to the total dissociation of some planets 
\citep{Etangs-2004,Baraffe-2004}. Conversely, cooling in the atmosphere may well lead to a much more limited mass loss close to the
lower limit derived by Vidal-Madjar et al. 
\citep{Yelle-2004,Hubbard-2006}.

Is a catastrophic fate a possible explanation for the  lack of planets with periods shorter that 3 days? The
existence of the VHJ is an important constraint to this type of scenario. Evaporation scenarios must be  able
to explain the observed evaporation rate of HD 209458b, while at the same  time accounting for the long-term
survival of the VHJ on even closer orbits. One possibility is that VHJ are not viable in the long term and
are fated to  catastrophic evaporation. In this case the three observed VHJ would simply be  too young to have
evaporated yet.

The aim of the present article is to examine the likelihood of the hypothesis 
that the three VHJ were found around parent stars that are very young compared 
to typical planet host stars and to the time scales of evaporation. For this 
purpose, three indicators of young stellar ages are combined: i) chromospheric
activity  \citep[e.g.][]{Henry-1996} ii) Lithium abundances and iii)
rotational velocity. 

The manuscript is organized as follows. Our observations and data reduction are
described in Sec. 2. In Sec. 3
our index $S_{US}$ is built and calibrated with 
respect to the index $S$ of Mount Wilson $(S_{MW})$. In Sec. 4, we derive 
the Li abundances for the observed stars. 
Projected rotational velocities available in the literature for stars hosting 
transiting planets
are compiled in Sec. 5. Using the different observables we derive ages in Sec. 6 
and 
discuss the implications of the derived ages for the planetary evolution under 
evaporation in Sec. 7. 


\section{Observations}

For 5 of the transiting planet host stars (OGLE-TR-10, 56, 111, 113 and TrES-1) 
the observations were carried out with 
the UVES spectrograph at the VLT-UT2 Kueyen 
telescope (program ID\,75.C-0185), between April and May 2005 (in service mode). 
The Dichroic \#1 390+580 mode was used. The red arm spectra cover the wavelength 
domain 
between 4780 and 6805\AA, with a gap between 5730 and 5835\AA, whereas the data 
obtained with blue arm cover the wavelength domain 
between 3260 and 4450\AA. For all stars, we adopted slit
widths of 0.9 arcsec and 1.1 arcsec, which provides a spectral resolution 
($\lambda$/$\Delta\lambda$) of the order of 50\,000 and 40\,000 in the red and 
blue arm, 
respectively. 

For each exposure, both the blue arm CCD and the red mosaic were read in 2x2 
bins 
to reduce the readout noise and increase the number of counts in each bin. This 
procedure 
does not compromise the resolving power, since the sampling of the CCD is still
higher (by a factor of 2) than the instrumental PSF.
For the brighter TrES-1, we choose a 1x1 binning for the red arm CCD.

\begin{figure}
\resizebox{\hsize}{!}{\includegraphics[angle=-90]{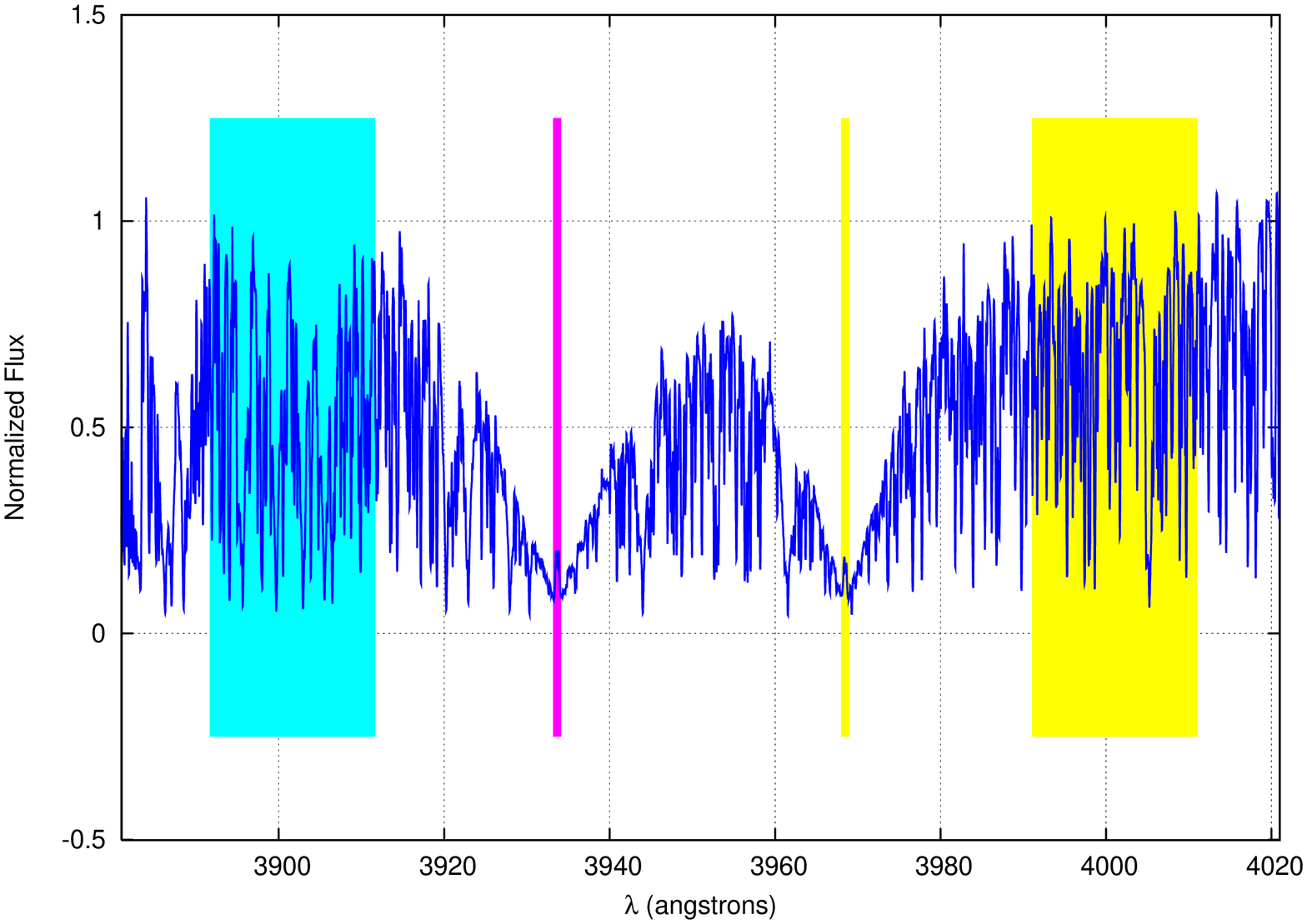}}
\resizebox{\hsize}{!}{\includegraphics[angle=-90]{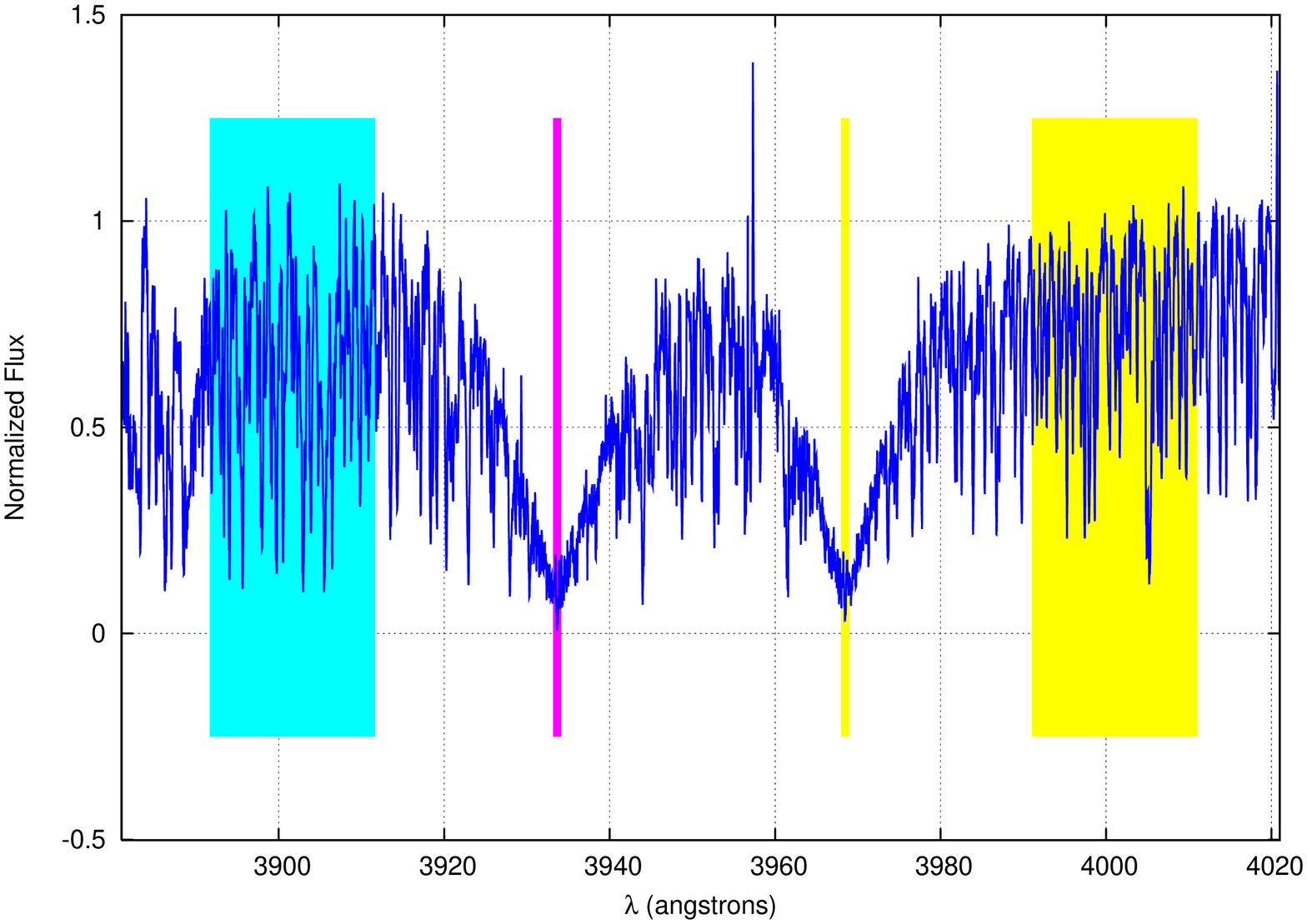}}
\caption{On line and continuum filters used in the computation of $S_{US}$. The 
UVES spectrum for TrES-1 (top panel) and OGLE-TR-10 are shown in the figure.}
\label{fig:filters}
\end{figure}

For the OGLE stars, particular attention was given to the orientation of 
the slit given the relative crowdedness of the fields. The angle was chosen
in each case using the images available at the OGLE 
website\footnote{http://www.astrouw.edu.pl/~ftp/ogle/index.html}, 
in order that no other star was present in the UVES slit.

Data reduction was carried out by the ESO staff using the UVES pipeline as part 
of 
the service mode data package.

Finally, for the analysis of HD189733 we have used the same CORALIE spectrum taken 
by \cite{Bouchy-2005b}. 

\section{Chromospheric Activity}

Chromospheric ages are commonly derived using the age-chromospheric activity  relation  constructed based on
the Mont Wilson (MW) chromospheric flux index  ($S_{MW}$)\citep[e.g.][]{Vaughan-1978,Henry-1996}. In order to
use MW calibration, one needs first to build a chromospheric  flux indicator (here called $S_{US}$) and then
calibrate it with respect to the  MW system ($S_{MW}$). This procedure has been carried out many times in
different  papers \citep[e.g.][]{Santos-2000a,Wright-2004,Saffe-2005}. Here we follow the  prescription  of
\cite{Santos-2000a}. 

The $S_{US}$ index is defined as $(F_H+F_K)/(C_H+C_K)$ where $F_H$ and $F_K$ are  the sum of the fluxes within
a 1-\AA\ box centered on the Ca H and K lines,  respectively. $C_H$ and $C_K$ are the total flux computed as
for $F_H$ and $F_K$  but for 20-\AA\ boxes centered at 4001.067\AA\ and 3901.67\AA\ , respectively.  In order to
calibrate $S_{US}$ as a function of $S_{MW}$ a number of common  stars  in both samples is needed. Our original
sample observed with UVES contains only  spectra of stars hosting transiting planets, therefore external data
for calibrators is needed. HARPS spectra, kindly provided by the HARPS team, were used  as calibrators.
Therefore, as a prior step before computing the flux index, the  HARPS  spectra were smoothed to match the
UVES blue-arm resolution of 40000 by  convolving a gaussian profile with the appropriate width and the UVES
data were rebinned to match the HARPS spectral sampling of 0.01\AA\ . 

Spectra of calibrators and planet host stars were brought to the rest  frame using the radial
velocities given by the cross-correlation function fit.  The  data were then trimmed from 3870\AA\ up to 4070\AA\
and normalized by a straight  line within this interval. Finally the $S_{US}$ is computed by adding the
relative intensity of the pixels within each filter.  
Figure~\ref{fig:filters} shows the spectral region used to compute $S_{US}$.

One might inquire whether a calibration built with data collected with one  instrument  can be used to
calibrate a different one. The index $S$ simply measures the  ratio between  the flux in the core of the Ca II
H and K lines with respect to a continuum  region next to  these lines. Therefore provided that i) the
instrumental responses are correctly  subtracted  within the region containing the Ca II lines and the
continuum, ii) the fluxes  are computed  exactly in the same way for both instruments and, iii) data have the
same  sampling  and resolution, the indices S will reflect solely the stellar contribution  regardless  of the
instrument used to collect the data.

Most of the instrumental contribution (blaze function, pixel-to-pixel  variations,  CCD response, etc.) is
removed by the division of the extracted spectrum by  the  flat field spectrum. The remaining effects are
subtracted by normalizing the  reduced  spectrum by a pseudo-continuum. Since the position of the true
continuum is not  known,  the normalization process will certainly introduce errors in our computed $S$  and 
therefore in the final ages. In order to quantify how changes in the continuum  are  reflected in the computed
$S_{US}$, we compared the $S_{US}$ values determined using  the normalized and non-normalized spectra. In the
worst cases the variations in  the $S_{US}$ were of 5\%. This error is much smaller than the variations  of
$S_{MW}$ which are typically of 20\%.

As an additional check of the validity of our $S_{US}-S_{MW}$ calibration, we  have compared the $S_{MW}$
derived using spectra collected with HARPS with those computed using  spectra from the UVES POP spectral library
\citep{Bagnulo-2004} for 6 stars in  common. For 5 stars whose $S_{MW}$ is below 0.2, the differences in the
$S_{MW}$ are  below 6\%. For the active star HD~22049 ($S_{MW}=0.467$ computed  from our $S_{US}$ using
Eq.~\ref{eq:fit}) the relative difference is 15\% which is largely due to its  intrinsic variability.  As
discussed in Sec. 6.1, we have conservatively adopted an error on $S_{MW}$ of  20\% when estimating our final
uncertainty on the derived ages.




\begin{table}
\begin{center}
\caption{$S_{MW}$ computed from
our chromospheric index flux $S_{US}$ using Eq.~\ref{eq:fit}
and the 
chromospheric fluxes corrected from the photospheric contribution ($<\log 
R^\prime_{HK}>$).
}
\label{table:ca}
\begin{tabular}{l@{\hspace{51.5pt}}ccc}
\hline\hline
Star & B-V & $S_{MW}$ & \multicolumn{1}{c}{$<\log R^\prime_{HK}>$} \\
\hline
TrES-1           & 0.848 & 0.247 &  -4.785 \\
OGLE-TR-10       & 0.606 & 0.192 &  -4.804 \\
OGLE-TR-56       & 0.589 & 0.120 &  -5.358 \\
OGLE-TR-111      & 0.933 & 0.275 &  -4.812 \\
OGLE-TR-113      & 1.020 & 0.448 &  -4.685 \\
OGLE-TR-132      & --    & --    & --      \\  
HD149026         & --    & --    & --      \\
HD189733	 & 0.898 & 0.437 &  -4.537 \\
HD209458         & 0.563 & 0.154$^a$ &  -4.988 \\
\hline
\end{tabular}
\end{center}
$^a$Taken from \cite{Wright-2004}
\end{table}

Using the $S_{US}$ computed as above and the $S_{MW}$ given by \cite{Henry-1996} 
we carried out a least-squares linear fit to the data which yields the relation:

\begin{equation}
S_{US}=0.06111\times S_{MW}-0.00341
\label{eq:fit}
\end{equation}

\begin{figure}
\resizebox{\hsize}{!}{\includegraphics[angle=-90]{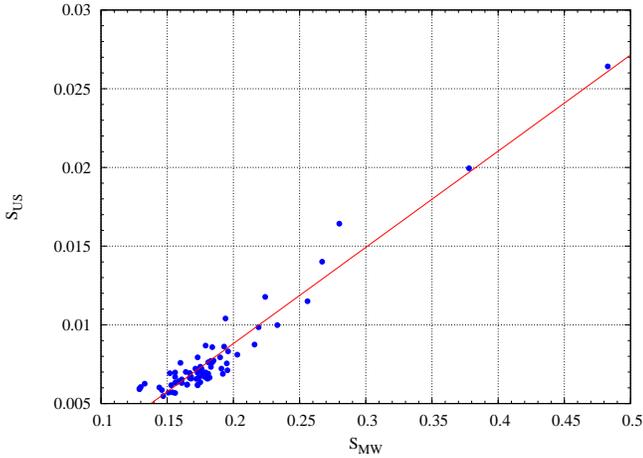}}
\caption{$S_{US}$ compared to the $S_{MW}$ values for a set of calibrators. 
The solid line is the best least-squared linear fit.
}
\label{fig:calib}
\end{figure}

Figure~\ref{fig:calib} shows our best-fit. The RMS of the residuals of the fit 
($\sigma_{fit}$) is  $8\times10^{-4}$. As pointed out by \cite{Santos-2000a}, 
$\sigma_{fit}$ includes not only uncertainties in the calibration procedure but
also intrinsic stellar variation. Eq.~\ref{eq:fit} can now be used to compute 
the $S_{MW}$. 

The Ca II H and K flux corrected for the photospheric  flux ($R'_{HK}=R_{HK}-R_{phot}$) is computed
using \cite{Noyes-1984}  prescription. The derived $S_{MW}$ yielded by Eq.~\ref{eq:fit}  and the corrected
chromospheric flux $R'_{HK}$ are given in  Table~\ref{table:ca}. We note that in this procedure for the OGLE
stars and TrES-1 we have used B-V  colors that were derived by inverting the B-V vs. (Teff,[Fe/H]) calibration
presented in  \citet[][]{Santos-2004b}.


No spectra in the CaII-line regions were available for HD149026 and OGLE-TR-132,
and no values for the S index are available in the literature for these stars 
either.
For HD209458 we used the value $S_{MW}=0.154$ derived by \cite{Wright-2004}, the 
effective temperature
 and metallicity of \cite{Santos-2004b} to compute the $R'_{HK}$ given in 
Table~\ref{table:ca}.

\section{Li abundances}

Lithium abundances were obtained as in \cite{Israelian-2004},
from a LTE analysis using a recent version of the radiative 
transfer code MOOG \citep{Sneden-1973} and a grid of \cite{Kurucz-1993} 
plane-parallel model atmospheres.
Equivalent widths (EWs) for the Li line near 6707.8\AA\ (or 
an upper limit for these) were measured using the IRAF 
{\tt splot} routine inside the {\tt echelle} package. This procedure 
gives good results since in our spectra the Li line was never 
strongly blended with any other line in the same spectral region.

The measured EWs (or upper limits for these) were then used to derive 
the Li abundances. The results are presented in Table\,\ref{table:li}.
We used the stellar parameters listed in \cite{Santos-2006}, 
also listed in Table\,\ref{table:li} for completeness. 
For more details on the derivation of Li abundances we refer the 
reader to \cite{Israelian-2004}.

For HD209458 we have taken the Li abundance from the study of \cite{Israelian-2004}. 
No Li abundances are available for both HD149026 and OGLE-TR-132.

\begin{table*}[t!]
\begin{center}
\caption{Stellar parameters and Li abundances for all known transiting planets.
See text for references.}
\label{table:li}
\begin{tabular}{l@{\hspace{5.5pt}}c@{\hspace{5.5pt}}ccrlcl}
\hline\hline
Star & T$_{eff}$ & $\log g$  & $\xi_{\mathrm{t}}$  & \multicolumn{1}{c}{[Fe/H]} 
&
\multicolumn{1}{c}{Reference}&\multicolumn{1}{c}{$A(Li)$}&\multicolumn{1}{c}{Reference} \\
& (K) & (c.g.s)&[km\,s$^{-1}$]&&\multicolumn{1}{c}{Stellar 
Parameters}&&\multicolumn{1}{c}{$A(Li)$}\\
\hline
TrES-1           & 5226$\pm$38  & 4.40$\pm$0.10 & 0.90$\pm$0.05 & 0.06$\pm$0.05    
&\cite{Santos-2006}	      &$<0.5$       & This paper\\
OGLE-TR-10       & 6075$\pm$86  & 4.54$\pm$0.15 & 1.45$\pm$0.14 & 0.28$\pm$0.10    
&\cite{Santos-2006}        &2.3$\pm$0.1  & This paper\\
OGLE-TR-56       & 6119$\pm$62  & 4.21$\pm$0.19 & 1.48$\pm$0.11 & 0.25$\pm$0.08    
&\cite{Santos-2006}        &2.7$\pm$0.1  & This paper\\
OGLE-TR-111      & 5044$\pm$83  & 4.51$\pm$0.36 & 1.14$\pm$0.10 & 0.19$\pm$0.07    
&\cite{Santos-2006}        &$<0.5$	    & This paper\\
OGLE-TR-113      & 4804$\pm$106 & 4.52$\pm$0.26 & 0.90$\pm$0.18 & 0.15$\pm$0.10    
&\cite{Santos-2006}        &$<0.2$	    & This paper\\
OGLE-TR-132      & 6411$\pm$179 & 4.86$\pm$0.14 & 1.46$\pm$0.36 & 0.43$\pm$0.18    
&\cite{Bouchy-2004}	      & --          &\\
HD149026         & 6147$\pm$50  & 4.26$\pm$0.07 & --            & 0.36$\pm$0.05    
&\cite{Sato-2005}	      &  --         &\\
HD189733	 & 5050$\pm$50  & 4.53$\pm$0.14 & 0.95$\pm$0.07 & $-$0.03$\pm$0.04 
&This paper	              & $<-0.1$     &This paper\\
HD209458         & 6117$\pm$26  & 4.48$\pm$0.08 & 1.40$\pm$0.06 &  0.02$\pm$0.03   
&\cite{Santos-2004b}     &2.7$\pm$0.1  &\cite{Israelian-2004}\\
\hline
\end{tabular}
\end{center}
\end{table*}

\section{Projected rotational velocity}

For most of our candidates the projected rotational velocities and the respective errors listed in
Table\,\ref{table:rot} were taken from the literature
\citep[][]{Queloz-2000,Pont-2004,Bouchy-2004,Bouchy-2005a,Sato-2005,Laughlin-2005}.  For OGLE-TR-10 and
OGLE-TR-56 the projected rotational velocities were derived  from HARPS spectra  using a calibration based on
the Cross-Correlation Function width \citep[see  e.g.][]{Santos-2002a}. 


\begin{table}
\begin{center}
\caption{Estimated $V\sin i$ for the transiting planet host stars studied in 
this paper.
}
\label{table:rot}
\begin{tabular}{lcl}
\hline\hline
Star & $V_{rot}$ (km s$^{-1}$) & Reference\\
\hline
TrES-1           & 1.08$\pm$0.3  & \citet[][]{Laughlin-2005} \\
OGLE-TR-10       & 7.0$\pm$1.0   & HARPS CCF\\
OGLE-TR-56       & 3.2$\pm$1.0   & HARPS CCF\\
OGLE-TR-111      & $<$5  & \citet[][]{Pont-2004}\\
OGLE-TR-113      & $<$5  & \citet[][]{Bouchy-2004}\\
OGLE-TR-132      & $<$5  & \citet[][]{Bouchy-2004}\\
HD149026         & 6.0$\pm$0.5   & \citet[][]{Sato-2005}\\
HD189733	 & 3.5$\pm$1.0   & \citet[][]{Bouchy-2005a}\\
HD209458         & 3.75$\pm$1.25  & \citet[][]{Queloz-2000}\\
\hline
\end{tabular}
\end{center}
\end{table}

We note that in our case the $V\sin i$ values correspond to the real equatorial rotational velocity, as long as
the rotational axis of the star is perpendicular  to the orbital plane of the transiting planet. This
hypothesis is supported by  observations of the Rossiter-McLauglin effect on HD209458 and HD189733
\citep[][]{Queloz-2000,Bouchy-2005a}.


\section{Age of stars hosting Transiting planets}


\subsection{Chromospheric ages}

\begin{figure}
\resizebox{\hsize}{!}{\includegraphics{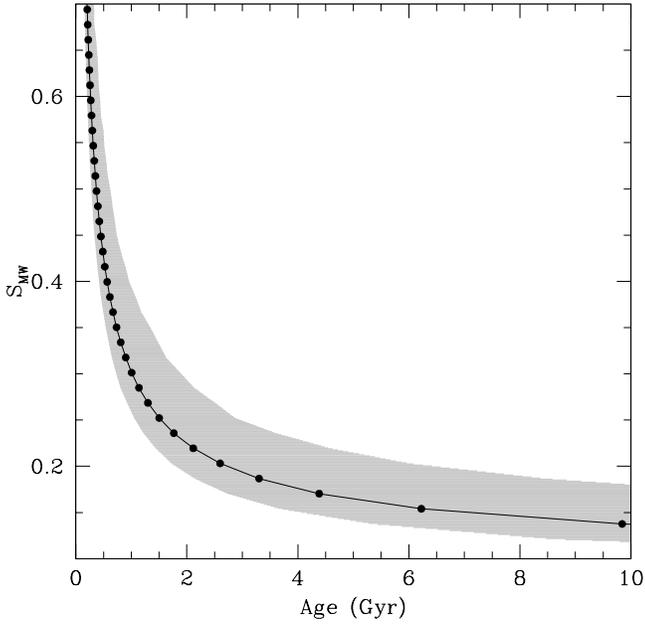}}
\caption{The projection of the age-activity relation for a $B-V=0.63 $ star
is shown by the circles and solid line. The shaded area around the solid line
is the 95\% confidence region for the true stellar age computed
assuming a 20\% variability level on the $S_{US}$. 
Due to the the flatness of the  activity-age calibration it is clearly seen that after 2-3Gyr
only a minimum age can be computed by this method.
}
\label{fig:mc}
\end{figure}

The formulae used in deriving stellar ages from chromospheric 
activity indices are summarized in \cite{Wright-2004}. A typical projection of 
the age-activity relation is shown in Fig.~\ref{fig:mc}. The relation between 
activity and age is steep for young ages and gets flatter for older ages. 
Additionally, stellar activity follows long-term cycles, and the use of 
instantaneous chromospheric flux instead of the flux averaged over an entire 
magnetic cycle leads to an over or under estimation of age depending on which 
moment of its cycle the star is observed. The classical example being the case 
of the Sun whose $\log R^\prime_{HK}$ varied from -4.75 to -5.10 
during the "Maunder Minimum'', corresponding to ages of 8.0 and 2.2 Gyr, 
respectively \citep[e.g.][]{Henry-1996}. 

\cite{Pace-2004} have studied empirically the reliability of 
chromospheric ages by measuring the chromospheric activity level in dwarfs 
belonging to 5 different open clusters with ages spanning
from 0.6 Gyr up to 4.5 Gyr. They found that after a  strong  decrease in 
chromospheric activity taking place 
in solar-type stars between the Hyades and the IC 4651 age (i.e., 0.6~Gyr and 
1.7~Gyr),
activity remains virtually constant for more than
3 Gyr. Hence an activity-age relation holds up to only about 2.0 Gyr.
Therefore, as cautioned by many authors, chromospheric activity indices are good 
indicators of youth, but cannot yield reliable age estimates beyond $\sim 2-3$ Gyr - 
as illustrated graphically on Fig.\ref{fig:mc}.

Consequently, since all stars in our sample show low levels of activity, we 
only derive lower age limits rather than specific age estimates from 
chromospheric activity. We used the $R'_{HK}$ values listed in 
Table\,\ref{table:ca} and the calibration of \cite{Henry-1996}. We calculated 
the 2-sigma lower limits on the ages by including as far as possible the sources 
of error on all intermediate steps between the computation of the $S_{US}$ and 
the final ages and allowed for an intrinsic dispersion of 20\% for the activity 
level of stars with identical mass and age.


\subsection{Li ages}

Li ages given in Table~\ref{table:ages} were derived by comparison with 
abundance curves of Li
as a function of effective temperature for clusters of different ages 
\citep{Jeffries-2002,Sestito-2005}.
Due to its fragility, Li is quickly destroyed during the pre-main sequence 
inside solar-mass stars which are
(almost) fully convective at the beginning of their lives. Li depletion also 
occurs on the main-sequence
although its causes are still under strong debate \citep[see review 
by][]{Deliyannis-1998}. For this reason Li
is widely used a youth indicator \cite[e.g.][]{Martin-1997}. 

For stars hotter than 6000 K (i.e., earlier than
G), the convective zones are not deep enough to bring Li to warmer regions thus 
Li is not depleted \citep[with
exception of the Li gap; ][]{Boesgaard-1986}. Hence Li abundance cannot be used 
as an age
indicator. This is the case for OGLE-TR-10 and OGLE-TR-56. 

The recent results of \cite{Sestito-2005} seem to indicate that Li-depletion  
occurs by quick
episodes rather than at a monotonic pace. This could imply that Li abundances 
can only yield lower limits for
the ages.


\begin{table}[b]
\begin{center}
\caption{Age constraints yielded by each method.
The 95\%-confidence interval (2-sigma)
for the chromospheric age computed using the mean value of $\log R^\prime_{HK}$ is 
shown along with 
and the age constraint from Li abundances. Whenever available age estimates 
found in the literature are also given.}
\label{table:ages}
\begin{tabular}{l@{\hspace{5.5pt}}c@{\hspace{5.5pt}}c@{\hspace{5.5pt}}c@{\hspace
{5.5pt}}c}
\hline\hline
Star 		 & age$_{CaII}$ 		& age$_{Li}$  & age   & Ref \\
		 & $2-\sigma$ lower lim.&	      &	other	      & other\\
\hline
TrES-1           &$>$ 1.1  & $>$0.6    &$2.5\pm1$&1\\
OGLE-TR-10       &$>$ 1.1  & --      &&\\
OGLE-TR-56       &$>$ 2  & --      &$3\pm1$&2\\
OGLE-TR-111      &$>$ 1.1  & $>$0.6  &&\\
OGLE-TR-113      &$>$ 0.7  & $>$0.6  &&\\
OGLE-TR-132      & --  & --      &&\\
HD149026         & --  & --      &$2\pm0.8$&3\\
HD189733         &$>$ 0.5  & $>$0.6      &&\\
HD209458         &$>$ 2  & -      &4.5&4\\
\hline
\end{tabular}
\end{center}
1-\cite{Sozzetti-2004b}; 2--\cite{Sasselov-2003}; \\
3--\cite{Sato-2005}; 4--\cite{Mazeh-2000}
\end{table}

\subsection{$V\sin i$ ages}


During the pre-main sequence phase,
angular momentum evolution is mainly driven by the magnetic coupling between 
the star and its disk 
\citep[e.g.][]{Koenigl-1991}. 
Due to differences in the disk-locking time-scales, solar type stars arrive on 
the zero age main 
sequence presenting a large spread in their rotation rates. From this point on,
any angular momentum evolution is dictated by stellar winds 
whose intensity is itself a function of magnetic activity and rotation 
\citep[e.g.][]{Kawaler-1988}. 
This creates a regulation mechanism leading to a steady decrease and to the 
eventual convergence 
of the rotation rate at about 1Gyr.

Therefore the use the rotation rate as a qualitative age diagnostic is limited. 
Nevertheless, 
comparing our $V_{rot}$ given in Table~\ref{table:rot} with Figure 2 of 
\cite{Bouvier-1997} we
see that despite the large dispersion in the points, the F and G stars with 
$V_{rot}<5$ km s$^{-1}$ are 
likely to be older than the Hyades ($\sim 600$ Myr), 
whereas for the cooler stars a $V_{rot} $ of 5 km s$^{-1}$ is still compatible 
with the age of the 
Pleiades ($\sim 100$ Myr). Similar conclusions can be drawn for OGLE-TR-10 or 
HD149026 whose
$V_{rot}=7$ and 6 km s$^{-1}$, respectively, can still be found (although 
unlikely) among the Pleiades members. 

The utility of the $V_{rot}$ values to derive stellar ages is limited, but
our results show that globally speaking our sample stars, and in particular 
those orbited
by VHJ, are very unlikely to be younger than the Pleiades \citep[see Figure 2 
of][]{Bouvier-1997}. 


A summary of the age constraints yielded by each method is given in 
Table~\ref{table:ages}.
As dicussed in Sec. 6.1, any attempt to assign a precise age based on the chromospheric activity is
meaningless due to the flatness of the age-activity calibration beyond 2-3Gyr.
Therefore, only the lower limit (2-sigma level) from chromospheric activity
and the age constraint from Li abundances are given rather than a precise age.
Age estimates found in the 
literature, quoted in Table~\ref{table:ages}, are consistent with these values. 

\section{Constraining planetary evaporation rates from stellar ages?}

%
%

 
The lower limits given in Table 4 allow us to state that the stars
harboring transiting planets studied here are not younger than 0.5 Gyr
and for most of them not younger than 1 Gyr at a 2-sigma level. Is
this compatible with a very efficient evaporation of these objects?

Efficient evaporation scenarios
\citep{Lammer-2003,Etangs-2004,Baraffe-2004}
imply that a fraction of the close-in planets do not survive, at least not as
gaseous giants. Therefore those planets seen in present-day  transit surveys correspond to the lucky
ones that were initially massive enough to escape (so far) evaporation. 
Our age determinations allow us in
principle to constrain the magnitude of the evaporation: 
stars harboring very close-in planets (and in particular the
transiting planets discovered so far) should be on average younger if
an efficient evaporation indeed took place. For very large evaporation
rates, old stars should have lost their close-in planetary companions,
except for the rare ones with very large initial masses.

{\bf In order to estimate how the mean age of stars harboring very close-in
planets depends on the magnitude of the planetary mass-loss, we
proceed as follows.
First, it is assumed that the initial masses of the planets are set
according to a fixed planetary mass function (see below). Monte-Carlo
approach is then used to draw a large number of star-planet systems
for which we calculate the fraction of planets that have evaporated away
as a function of stellar age and evaporation rate.
By re-normalizing this fraction we
derive a distribution of the age of stars with planets as a function
of the assumed planetary evaporation rate. 

The process described above is carried out assuming the following hypotheses:
}
\begin{enumerate}

\item {\bf The initial planetary mass function for very-close in planets is
  the same as for planets observed by radial velocimetry with periods
  smaller than 365 days (a value chosen to avoid biases for planets
  with masses $> 0.5\rm\,M_{Jup}$, and which is found to affect  the results only
  moderately). The sample used is taken from
  J. Schneider's web page (www.obspm.fr/planets) and contains 99
  planets. }

\item Planets disappear from the sample when their mass becomes
  smaller than 0.5\,M$_{\rm Jup}$, either because of a runaway
  evaporation \citep{Baraffe-2004}, or because the presence of a core
  makes them too small to be detected by the transit method
  \citep[e.g.][]{Guillot-2005}.

\item The stellar formation rate (including stars with planets) has
  been constant for the past 10 Gyr. 
\end{enumerate}

With the present understanding of planet formation and migration, our
first hypothesis should be conservative (i.e. allow for higher
evaporation rates), as more massive planets are expected to migrate
less rapidly and efficiently to become HJ or VHJ planets
\citep[e.g.]{Udry-2003,Ida-2004a}. This is also the case of the third
hypothesis, as the progressive enrichment of the Galaxy in metals by
stellar nucleosynthesis implies that stars with planets should be on
average younger. More detailed studies are needed to address more
precisely these points and assess their effect on the results.

\begin{figure}
\resizebox{\hsize}{!}{\includegraphics{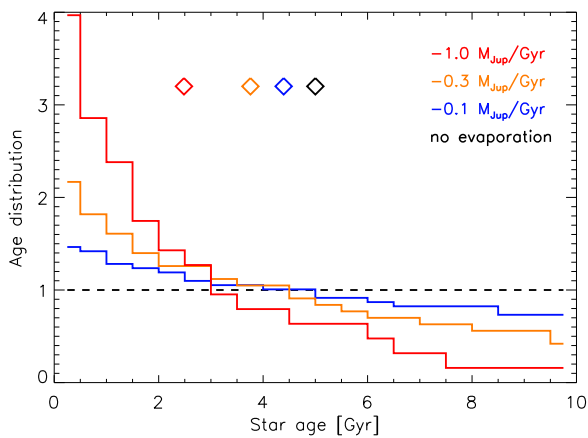}}
\caption{Age distribution of stars with planets, as a function of the
  planetary evaporation rate. We assume that the unperturbed sample of
  stars with planets have a uniform age distribution between 0 and 10
  Gyr (dashed line). The progressively steeper distributions
  correspond to evaporation rates of 0.1 (blue), 0.3 (orange) and
  1.0\,M$_{\rm Jup}$/Gyr (red), respectively. The diamonds represent
  the mean ages of the different distributions.}
\label{fig:prob}
\end{figure}

Figure~\ref{fig:prob} shows the resulting age distribution of stars
with planets for different planetary evaporation rates. For the most
extreme case of a 1\,M$_{\rm Jup}$/Gyr mass loss, the mean of the age
distribution is lowered to only 2.5 Gyr. For comparison, according to
\citep{Baraffe-2004}, the (maximal) evaporation rates derived from an
energy-limited approach is almost constant in time and range from
0.1\,M$_{\rm Jup}$/Gyr (HD209458b, OGLE-TR-10b) to almost 0.7\,M$_{\rm
Jup}$/Gyr (OGLE-TR-132b).

{\bf In addition to the age distribution derived in 
Figure~\ref{fig:prob}, one can now check whether the sample of stars with age
determinations derived in this paper 
is consistent with an energy-limited evaporation
scenario.
Using again a Monte-Carlo approach and adopting 
the evaporation rates derived by
\citet{Baraffe-2004},  the distribution of possible ages for
each star with transiting planet studied is derived.
We then compute the number
of times that these "Monte-Carlo" stellar ages are compatible with the inferred
minimal ages, both in the energy-limited evaporation scenario and in
the no-evaporation scenario. \emph{We found that the latter was between 2
and 3 times more likely, i.e. that the energy-limited evaporation
scenario was, in this case, marginally inconsistent with the
observations}.

Unfortunately, an additional limitation
has to be considered, namely,  the
fact that stars with young ages ($<0.5$\,Gyr) are quite active and as
such are unlikely to be detected as having a planet (both by the
transit method and by radial velocimetry). Due to this further constraint
and given the small sample used in this work, we cannot conclude in
favor or against the energy-limited evaporation scenario.
}

\section{Conclusions}

We have shown that the analysis of high resolution spectra of stars
with planets could be used to determine lower limits on their age
using Li abundances, the Ca~II activity-age relation and from the
analysis of the stellar rotational velocity. Applied to six stars
known to host transiting giant planets, OGLE-TR-10, 56, 111, 113,
TrES-1, and HD189733, we were able to show that none appears to be
younger than 1\,Gyr at the 2-sigma level. 

{\bf In principle, this could be used to constrain the amount of mass loss
experienced by the planets, since high evaporation rates should yield
smaller average ages for the stars with very close-in giant planets. 
Because of the various uncertainties and small size of the sample studied here,
the results remain inconclusive concerning
evaporation. However, it highlights the importance of a careful
determination of the ages of stars with planets, in particular for the
stars harboring very close-in planets (``Pegasids''), but also, as a
means of comparison, for stars with planets on orbits with
longer-periods. }

We hence recommend to pursue the fine determination of the ages of
stars with planets, in priority for the transiting planets as they are
progressively being detected \citep[for a direct
application see][]{Guillot-2006}, for non-transiting very close-in
giant planets (at less than 0.1 AU from their star), and then for a
sample of stars with planets that are orbiting at larger distances.

\begin{acknowledgements} 
Support from Funda\c{c}\~ao para a Ci\^encia e a Tecnologia 
(Portugal) to N.C.S. in the form of a scholarship (reference
SFRH/BPD/8116/2002) and a grant (reference POCI/CTE-AST/56453/2004) is
gratefully acknowledged. T.G. thanks the Programme National de
Plan\'etologie for support. 
  \end{acknowledgements}

\bibliographystyle{aa}
\bibliography{mybib}

\begin{thebibliography}{44}
\expandafter\ifx\csname natexlab\endcsname\relax\def\natexlab#1{#1}\fi

\bibitem[{{Bagnulo} {et~al.}(2003){Bagnulo}, {Jehin}, {Ledoux}, {Cabanac},
  {Melo}, {Gilmozzi}, \& {The ESO Paranal Science Operations
  Team}}]{Bagnulo-2004}
{Bagnulo}, S., {Jehin}, E., {Ledoux}, C., {et~al.} 2003, The Messenger, 114, 10

\bibitem[{{Baraffe} {et~al.}(2004){Baraffe}, {Selsis}, {Chabrier}, {Barman},
  {Allard}, {Hauschildt}, \& {Lammer}}]{Baraffe-2004}
{Baraffe}, I., {Selsis}, F., {Chabrier}, G., {et~al.} 2004, A\&A, 419, L13

\bibitem[{{Boesgaard} \& {Tripicco}(1986)}]{Boesgaard-1986}
{Boesgaard}, A.~M. \& {Tripicco}, M.~J. 1986, \apj, 303, 724

\bibitem[{{Bouchy} {et~al.}(2005{\natexlab{a}}){Bouchy}, {Pont}, {Melo},
  {Santos}, {Mayor}, {Queloz}, \& {Udry}}]{Bouchy-2005a}
{Bouchy}, F., {Pont}, F., {Melo}, C., {et~al.} 2005{\natexlab{a}}, A\&A, 431,
  1105

\bibitem[{{Bouchy} {et~al.}(2004){Bouchy}, {Pont}, {Santos}, {Melo}, {Mayor},
  {Queloz}, \& {Udry}}]{Bouchy-2004}
{Bouchy}, F., {Pont}, F., {Santos}, N.~C., {et~al.} 2004, A\&A, 421, L13

\bibitem[{{Bouchy} {et~al.}(2005{\natexlab{b}}){Bouchy}, {Udry}, {Mayor},
  {Moutou}, {Pont}, {Iribarne}, {Da Silva}, {Ilovaisky}, {Queloz}, {Santos},
  {Segransan}, \& {Zucker}}]{Bouchy-2005b}
{Bouchy}, F., {Udry}, S., {Mayor}, M., {et~al.} 2005{\natexlab{b}}, A\&A, in
  press

\bibitem[{{Bouvier}(1997)}]{Bouvier-1997}
{Bouvier}, J. 1997, Memorie della Societa Astronomica Italiana, 68, 881

\bibitem[{{Deliyannis} {et~al.}(2000){Deliyannis}, {Pinsonneault}, \&
  {Charbonnel}}]{Deliyannis-1998}
{Deliyannis}, C.~P., {Pinsonneault}, M.~H., \& {Charbonnel}, C. 2000, in IAU
  Symposium 198, 61

\bibitem[{{Gaudi} {et~al.}(2005){Gaudi}, {Seager}, \&
  {Mallen-Ornelas}}]{Gaudi-2005}
{Gaudi}, B.~S., {Seager}, S., \& {Mallen-Ornelas}, G. 2005, ApJ, 623, 472

\bibitem[{{Guillot}(2005)}]{Guillot-2005}
{Guillot}, T. 2005, Annual Review of Earth and Planetary Sciences, 33, 493

\bibitem[{{Guillot} {et~al.}(2006){Guillot}, {Santos}, {Pont}, {Iro}, {Melo},
  \& {Ribas}}]{Guillot-2006}
{Guillot}, T., {Santos}, N.~C., {Pont}, F., {et~al.} 2006, \aap, 453, L21

\bibitem[{{Henry} {et~al.}(1996){Henry}, {Soderblom}, {Donahue}, \&
  {Baliunas}}]{Henry-1996}
{Henry}, T.~J., {Soderblom}, D.~R., {Donahue}, R.~A., \& {Baliunas}, S.~L.
  1996, \aj, 111, 439

\bibitem[{{Hubbard} {et~al.}(2005){Hubbard}, {Hattori}, {Burrows}, {Hubeny}, \&
  {Sudarsky}}]{Hubbard-2006}
{Hubbard}, W.~B., {Hattori}, M.~F., {Burrows}, A., {Hubeny}, I., \& {Sudarsky},
  D. 2005, ArXiv Astrophysics e-prints

\bibitem[{{Ida} \& {Lin}(2004)}]{Ida-2004a}
{Ida}, S. \& {Lin}, D.~N.~C. 2004, ApJ, 604, 388

\bibitem[{{Israelian} {et~al.}(2004){Israelian}, {Santos}, {Mayor}, \&
  {Rebolo}}]{Israelian-2004}
{Israelian}, G., {Santos}, N.~C., {Mayor}, M., \& {Rebolo}, R. 2004, A\&A, 414,
  601

\bibitem[{{Jeffries} {et~al.}(2002){Jeffries}, {Totten}, {Harmer}, \&
  {Deliyannis}}]{Jeffries-2002}
{Jeffries}, R.~D., {Totten}, E.~J., {Harmer}, S., \& {Deliyannis}, C.~P. 2002,
  \mnras, 336, 1109

\bibitem[{{Kawaler}(1988)}]{Kawaler-1988}
{Kawaler}, S.~D. 1988, \apj, 333, 236

\bibitem[{{K\"onigl}(1991)}]{Koenigl-1991}
{K\"onigl}, A. 1991, \apjl, 370, L39

\bibitem[{{Kurucz}(1993)}]{Kurucz-1993}
{Kurucz}, R. 1993, ATLAS9 Stellar Atmosphere Programs and 2 km/s grid.~Kurucz
  CD-ROM No.~13.~ Cambridge, Mass.: Smithsonian Astrophysical Observatory,
  1993., 13

\bibitem[{{Lammer} {et~al.}(2003){Lammer}, {Selsis}, {Ribas}, {Guinan},
  {Bauer}, \& {Weiss}}]{Lammer-2003}
{Lammer}, H., {Selsis}, F., {Ribas}, I., {et~al.} 2003, \apjl, 598, L121

\bibitem[{{Laughlin} {et~al.}(2005){Laughlin}, {Wolf}, {Vanmunster},
  {Bodenheimer}, {Fischer}, {Marcy}, {Butler}, \& {Vogt}}]{Laughlin-2005}
{Laughlin}, G., {Wolf}, A., {Vanmunster}, T., {et~al.} 2005, \apj, 621, 1072

\bibitem[{{Lecavelier des Etangs} {et~al.}(2004){Lecavelier des Etangs},
  {Vidal-Madjar}, {McConnell}, \& {H{\' e}brard}}]{Etangs-2004}
{Lecavelier des Etangs}, A., {Vidal-Madjar}, A., {McConnell}, J.~C., \& {H{\'
  e}brard}, G. 2004, A\&A, 418, L1

\bibitem[{{Martin}(1997)}]{Martin-1997}
{Martin}, E.~L. 1997, \aap, 321, 492

\bibitem[{{Mayor} \& {Queloz}(1995)}]{Mayor-1995}
{Mayor}, M. \& {Queloz}, D. 1995, Nature, 378, 355

\bibitem[{{Mazeh} {et~al.}(2000){Mazeh}, {Naef}, {Torres}, {Latham}, {Mayor},
  {Beuzit}, {Brown}, {Buchhave}, {Burnet}, {Carney}, {Charbonneau}, {Drukier},
  {Laird}, {Pepe}, {Perrier}, {Queloz}, {Santos}, {Sivan}, {Udry}, \&
  {Zucker}}]{Mazeh-2000}
{Mazeh}, T., {Naef}, D., {Torres}, G., {et~al.} 2000, ApJ, 532, L55

\bibitem[{{Noyes} {et~al.}(1984){Noyes}, {Hartmann}, {Baliunas}, {Duncan}, \&
  {Vaughan}}]{Noyes-1984}
{Noyes}, R.~W., {Hartmann}, L.~W., {Baliunas}, S.~L., {Duncan}, D.~K., \&
  {Vaughan}, A.~H. 1984, \apj, 279, 763

\bibitem[{{Pace} \& {Pasquini}(2004)}]{Pace-2004}
{Pace}, G. \& {Pasquini}, L. 2004, \aap, 426, 1021

\bibitem[{{Pollack} {et~al.}(1996){Pollack}, {Hubickyj}, {Bodenheimer},
  {Lissauer}, {Podolak}, \& {Greenzweig}}]{Pollack-1996}
{Pollack}, J., {Hubickyj}, O., {Bodenheimer}, P., {et~al.} 1996, Icarus, 124,
  62

\bibitem[{{Pont} {et~al.}(2004){Pont}, {Bouchy}, {Queloz}, {Santos}, {Melo},
  {Mayor}, \& {Udry}}]{Pont-2004}
{Pont}, F., {Bouchy}, F., {Queloz}, D., {et~al.} 2004, A\&A, 426, L15

\bibitem[{{Queloz} {et~al.}(2000){Queloz}, {Mayor}, {Weber}, {Bl{\' e}cha},
  {Burnet}, {Confino}, {Naef}, {Pepe}, {Santos}, \& {Udry}}]{Queloz-2000}
{Queloz}, D., {Mayor}, M., {Weber}, L., {et~al.} 2000, A\&A, 354, 99

\bibitem[{{Saffe} {et~al.}(2005){Saffe}, {Gomez}, \& {Chavero}}]{Saffe-2005}
{Saffe}, C., {Gomez}, M., \& {Chavero}, C. 2005, ArXiv Astrophysics e-prints

\bibitem[{{Santos} {et~al.}(2004){Santos}, {Israelian}, \&
  {Mayor}}]{Santos-2004b}
{Santos}, N.~C., {Israelian}, G., \& {Mayor}, M. 2004, A\&A, 415, 1153

\bibitem[{{Santos} {et~al.}(2000){Santos}, {Mayor}, {Naef}, {Pepe}, {Queloz},
  {Udry}, \& {Blecha}}]{Santos-2000a}
{Santos}, N.~C., {Mayor}, M., {Naef}, D., {et~al.} 2000, A\&A, 361, 265

\bibitem[{{Santos} {et~al.}(2002){Santos}, {Mayor}, {Naef}, {Pepe}, {Queloz},
  {Udry}, {Burnet}, {Clausen}, {Helt}, {Olsen}, \& {Pritchard}}]{Santos-2002a}
{Santos}, N.~C., {Mayor}, M., {Naef}, D., {et~al.} 2002, A\&A, 392, 215

\bibitem[{{Santos} {et~al.}(2006){Santos}, {Pont}, {Melo}, {Israelian},
  {Bouchy}, {Mayor}, {Moutou}, {Queloz}, {Udry}, \& {Guillot}}]{Santos-2006}
{Santos}, N.~C., {Pont}, F., {Melo}, C., {et~al.} 2006, A\&A, accepted

\bibitem[{{Sasselov}(2003)}]{Sasselov-2003}
{Sasselov}, D.~D. 2003, ApJ, 596, 1327

\bibitem[{{Sato} {et~al.}(2005){Sato}, {Fischer}, {Henry}, {Laughlin},
  {Butler}, {Marcy}, {Vogt}, {Bodenheimer}, {Ida}, {Toyota}, {Wolf}, {Valenti},
  {Boyd}, {Johnson}, {Wright}, {Ammons}, {Robinson}, {Strader}, {McCarthy},
  {Tah}, \& {Minniti}}]{Sato-2005}
{Sato}, B., {Fischer}, D.~A., {Henry}, G.~W., {et~al.} 2005, ArXiv Astrophysics
  e-prints

\bibitem[{{Sestito} \& {Randich}(2005)}]{Sestito-2005}
{Sestito}, P. \& {Randich}, S. 2005, \aap, 442, 615

\bibitem[{{Sneden}(1973)}]{Sneden-1973}
{Sneden}, C. 1973, Ph.D. Thesis, Univ. of Texas

\bibitem[{{Sozzetti} {et~al.}(2004){Sozzetti}, {Yong}, {Torres}, {Charbonneau},
  {Latham}, {Allende Prieto}, {Brown}, {Carney}, \& {Laird}}]{Sozzetti-2004b}
{Sozzetti}, A., {Yong}, D., {Torres}, G., {et~al.} 2004, ApJ, 616, L167

\bibitem[{{Udry} {et~al.}(2003){Udry}, {Mayor}, \& {Santos}}]{Udry-2003}
{Udry}, S., {Mayor}, M., \& {Santos}, N. 2003, A\&A, 407, 369

\bibitem[{{Vaughan} {et~al.}(1978){Vaughan}, {Preston}, \&
  {Wilson}}]{Vaughan-1978}
{Vaughan}, A.~H., {Preston}, G.~W., \& {Wilson}, O.~C. 1978, \pasp, 90, 267

\bibitem[{{Wright} {et~al.}(2004){Wright}, {Marcy}, {Butler}, \&
  {Vogt}}]{Wright-2004}
{Wright}, J.~T., {Marcy}, G.~W., {Butler}, R.~P., \& {Vogt}, S.~S. 2004, \apjs,
  152, 261

\bibitem[{{Yelle}(2004)}]{Yelle-2004}
{Yelle}, R.~V. 2004, Icarus, 170, 167

\end{thebibliography}

\end{document}